\documentclass[]{aa}  

\usepackage{graphicx}
\usepackage{txfonts}
\usepackage{lipsum}
\usepackage{subcaption}         
\usepackage{lscape}             
\usepackage{placeins}           

\usepackage[dvipsnames]{xcolor} 
\usepackage{multirow} 
\usepackage{multicol}
\usepackage{soul} 

\usepackage{enumitem}
\renewcommand\makeLineNumber{} 
                                
\usepackage[]{hyperref}
\usepackage{hyperref}
\hypersetup{
    colorlinks=true,
    linkcolor=blue,
    citecolor=blue,
    filecolor=magenta,      
    urlcolor=blue,
}

\newcommand{\parfrac}[2]{\frac{\partial #1}{\partial #2}}
\newcommand{\low}[1]{_\text{#1}}
\newcommand{\E}[1]{\times10^{#1}}

\begin{document}

   \title{The evolution and internal structure of Neptunes and sub-Neptunes}

    \subtitle{The importance of thermal conductivity in non-convective regions}

   \author{Mark Eberlein\inst{1}\thanks{Corresponding author: mark.eberlein@uzh.ch}
        \and Ravit Helled\inst{1}
        }

   \institute{Department of Astrophysics, University of Zurich, Winterthurerstrasse 190, 8057 Zurich, Switzerland}

   \date{Received 21 July 2025 / Accepted 3 September 2025}

  \abstract
   {Neptunes and sub-Neptunes are typically modeled under the assumption that the interior is adiabatic and consists of distinct layers. However, this assumption is oversimplified, as formation models indicate that composition gradients can exist. Such composition gradients can significantly affect the planetary thermal evolution by inhibiting convection. In non-convective layers, the heat transport is governed by multiple processes, and each is relevant in different regions within the planet.}
   {
   We investigate how the evolution and internal structure of Neptunes and sub-Neptunes is affected when considering non-convective layers and the sensitivity of the results on the assumed thermal conductivity. } 
   {We simulated the planetary evolution of such objects with an appropriate implementation for the conductivity by considering thermal transport via radiation, electrons, and vibrational conductivity. We considered planetary masses of 5, 10, and 15 M$_{\oplus}$; three different initial energy budgets; and two different primordial composition profiles.} 
   {
   We find that the assumed conductivity significantly affects the planetary thermal evolution. 
   We show that the commonly used conductivity assumption is inappropriate for modeling this planetary type. 
   Furthermore, we find that the inferred radii deviate  by $\sim20\%$ depending on the assumed conductivity. The uncertainty on the primordial entropy in planets with non-convective layers leads to a difference of $\sim25\%$ in the radii. This shows that the theoretical uncertainties are significantly larger than the observed ones and emphasizes the importance of these parameters. 
   }
   {We conclude that the characterization and modeling of intermediate-mass gaseous planets strongly depend on the modeling approach and the model assumptions. We demonstrate that the existence of composition gradients significantly affects the inferred radius. 
   We suggest that more data on thermal conductivities, particularly for partially ionized material and mixtures, as well as better constraints on the primordial thermal state of such planets are necessary.
   }


   \maketitle


\section{Introduction}
Intermediate-mass planets make up the majority of the detected exoplanet population. Their interiors are assumed to be composed of mostly rocks and ices surrounded by a hydrogen-helium (H-He) envelope. Unlike super-Earth planets, their radii are inflated by the presence of light elements \citep[e.g.][]{Rogers+2011,2018ApJ...866...49L}. The mass fraction of H-He strongly influences the radius of these planets regardless of the internal energy or the external stellar flux. 
Apart from the existence of an H-He atmosphere and a deep interior that is dominated by heavy elements, their detailed structure is highly uncertain. Constrained by only radius and mass, the composition profile of the detected exoplanet population is highly degenerate. Even for Uranus and Neptune, where gravitational data provide additional constraints on the planetary density distribution, the structure and bulk composition is still uncertain \citep[e.g.][]{Nettelmann+2013, NeunschwanderHelled2022, Morf+2024}.

Since the internal planetary structure can change as a planet evolves, static models may be further constrained by modeling planetary evolution. Often, planets are assumed to be fully convective, i.e., adiabatic, which imposes strong assumptions on both their evolution and structure. 
However, it recently became evident that assuming a homogeneous and adiabatic structure is an oversimplification. Composition gradients, which inhibit convection, can form due to the continuous accretion of solids and gas \citep[e.g.][]{VallettaHelled2022} or by accreting different materials when migrating over ice lines in the protoplanetary disk \citep[e.g.][]{SchneiderBitsch2021,Eberlein+2024}. 
Indeed, it has been suggested that both Uranus and Neptune have composition gradients in their deep interiors that can survive for billions of years \citep{VazanHelled2020, TejadaArevalo2025}. 
Composition gradients are not always primordial and can be caused by other processes, such as phase separation and the rainout of heavy elements. This further complicates the picture and can create (additional) composition gradients during the evolution of a planet \citep[e.g.,][]{Stevenson1980SaturnLuminosity, Lorenzen+2009, Redmer+2009, Puestow+2016, SchoettlerRedmer2018, Chang+2024, CanoAmoros+2024}. In addition, depending on the opacity, a deep radiative layer can also form within the planet \citep[e.g.,][]{2023A&A...680L...2H,MuellerHelled2024}.

In non-convective regions, the heat transport is governed by multiple thermal transport mechanisms -- radiative, diffusive, and conductive. Usually, the diffusive transport where hot particles diffuse into regions with colder temperatures is insignificant compared to radiation and conduction. In the outer parts of the planet, it is mostly the photons that contribute to thermal transport. Material emits and reabsorbs thermal radiation, and given a temperature gradient, this leads to a net diffusion of thermal energy that depends on the Rosseland-mean opacity \citep{Freedman+2014, MarigoAringer2009, Marigo+2024}. In denser regions, where the temperature is still sufficiently low so that the material is not ionized, conductivity is governed by the vibration of molecules, similar to phonons in solids. \citet{Ross+1984} reviewed the phonon model for metals applied to anharmonic non-metallic crystals. This model was used for the thermal conductivity of the Earth's core-mantle boundary \citep{Stamenkovic+2011}. More recently, \citet{French2019} used ab initio calculations of water to create an empirical fit of the thermal conductivity contribution from the nuclei. In the deep hot interior of the planet, free electrons enhance the thermal conductivity. \citet{Cassisi+2007} elaborated on the chemical picture of the electron conductivity by considering the density and scattering processes of electrons. \citet{FrenchRedmer2017} complemented the conductivities of the water nucleus by calculating the electron contribution.

In this paper, we investigate the influence of thermal conductivity on the evolution of Neptunes and sub-Neptunes.
We compare different cases for the conductivity to assess how it affects the radius evolution and the inferred internal structure.
Our paper is structured as follows. In Section \ref{sec:Methods}, we summarize the physical processes and the methods. In section \ref{sec:Results} we show the results of our study and discuss them in section \ref{sec:Discussion}. Finally, our summary and conclusions are presented in section \ref{sec:SummaryAndConclusion}.

\section{Methods}\label{sec:Methods}
We modeled the planetary evolution by solving the stellar structure equations \citep[e.g.][]{Kippenhahn+2013} using the Henyey method \citep{Henyey+1965} with the Modules for Experiments in Stellar Astrophysics (MESA) code \citep{Paxton+2011, Paxton+2013, Paxton+2015, Paxton+2018, Paxton+2019, Jermyn+2023}. We modified the code to make it applicable to medium-sized planets. 
We used the equation of state (EoS) presented by \citet{Mueller+2020a, Mueller+2020b, MuellerHelled2021, MuellerHelled2024} that combines the EoS with non-ideal interaction \citep{ChabrierDebras2021} for H-He with the heavy-element EoS for H$_2$O and SiO$_2$ \citep{More+1988, Vazan+2013} in a 50/50 rock-water mixture. 
The tables were implemented using a modified version of the MESA extension 'custom EoS' \citep{KnierimHelled2024, Helled+2025}. The modifications enabled the solver to utilize linear interpolations within the EoS tables rather than the standard bicubic spline interpolation scheme employed by MESA. 

Furthermore, we changed the radiative opacities and thermal conductivities as discussed below. We started the evolution with a given initial thermal state, composition profile, and mass as discussed in section \ref{sec:ModelSetup} and evolved the planet for 10 Gyr. 
The atmospheric cooling was simulated using the irradiated gray implementation in MESA based on \citet{GuillotHavel2011}. We adopted an equilibrium temperature of $T\low{eq}=400$ K and a mean visible-to-thermal opacity ratio of $\kappa\low{V}/\kappa\low{th}=0.03$ based on the fit provided by \citet{PoserRedmer2024}. Although our chosen equilibrium temperature lies outside the original fit range of 500–4000 K, the authors estimate the fit uncertainty to be approximately $\pm0.1$, which exceeds the difference in the opacity ratio between 400 K and 500 K.
We used the Ledoux criterion to determine if a region is convective or non-convective. Because of numerically challenging simulations, we did not change the composition over time.

\subsection{Thermal conductivity and opacity}\label{sec:ConductivityAndOpacity} 
A temperature gradient, $\nabla T$, within a planet creates a thermal flux, $F\low{thermal}$. 
The flux within the planetary interior can be modeled as a diffusive process such that it follows Fourier's law: 
\begin{equation}
    F\low{thermal} = - k\nabla T, 
\end{equation}
given the thermal conductivity, $k$. The total conductivity, $k\low{tot}$, which is calculated as 
\begin{equation}
    k\low{tot}=k\low{rad}+k\low{vib}+k\low{elec},
\end{equation}
is the sum of all processes, where $\kappa\low{rad}$ is the conductivity contribution from photons, while $k\low{vib}$ and $k\low{elec}$ are the conductivity contributions from vibrations in a dense fluid and from electrons, respectively.

The radiative conductivity can be converted to radiative opacity, $\kappa\low{rad}$, which measures how much light is absorbed in a medium using  
\begin{equation}
    \kappa\low{rad} = \frac{16\sigma}{3} \frac{T^3}{\rho k\low{rad}},
\end{equation}
where $\sigma$ is the Stefan-Boltzmann constant, $T$ the temperature, and $\rho$ the density. In the following, we use the term radiative conductivity to refer to thermal transport by photons.  
The standard low temperature opacity tables in MESA that cover high metallicities are an extended version of the Freedman opacities \citep{Freedman+2008, Freedman+2014}. They have been tabulated for a fixed solar H-He ratio and different metallicity fractions up to $Z=1$, relevant for planets. However, the reliable range stops at $\log (T/\text{K}) =3.8$ and significantly underestimates the opacity. In this region, MESA uses high temperature opacities instead, which do not cover high $Z\geq0.2$ materials. 
To improve on that, we created tables similar to the Freedman tables using the AESOPUS2.1 \citep{MarigoAringer2009, Marigo+2024} web interface. The recently updated method allowed us to the create tables in the range for temperatures up to $\log (T/\text{K}) =4.5$ with arbitrary values for the metallicity. 
The AESOPUS2.1 opacities were compared to the Freedman tables and to another set of opacities from \citet{Malygin+2014}. In the regime relevant for planets (high $\log R$), the AESOPUS2.1 opacities are generally higher than the opacities from the other two studies, especially below temperatures of $\log (T/\text{K}) \lesssim 3.3$ for \citet{Malygin+2014} and $\log (T/\text{K}) \lesssim 3.5$ for the Freedman tables. The differences might originate from a more complete list of molecular lines over the entire temperature region \citep{Marigo+2024}. As a result, the  AESOPUS2.1 opacities seem to be more reliable. We note that the latest release of MESA includes a set of opacity tables made with the AESOPUS2.1 code. However, these tables only include low metallicities and are therefore inappropriate for modeling a large range of planetary types.
(For details on how we created the tables and used them in MESA, see the Appendix \ref{sec:AESOPUSTable}.) 

For the vibrational conductivity, we used two models and compared them with each other. From the theory of phonons in metals, \citet{Ross+1984} reviewed the application to anharmonic non-metallic crystals. We used the formulation from \citet{Stamenkovic+2011} and took the reference conductivity for MgSiO$_3$ $k\low{vib, ref}(\rho\low{ref}=3.87\text{ g/cm}^3,\, T\low{ref}=2000\text{ K})=0.71$ W/m/K \citep[see Table 1][]{Stamenkovic+2011}. More details can be found in  Appendix \ref{sec:VibrationalConductivity}. 
The second model we used is based on the empirical fit to ab initio calculated conductivities of water (eq. 7 in French 2019). They provide an analytical function that is suitable for densities in the range of  $\rho\in(0.2, 10)$ g/cm$^3$ and temperatures of $T\in(600,\, 50\,000)$ K.

The electron contribution to the thermal conductivity in MESA is modeled by default with improved tables from Cassis et al. (2007). This conductivity assumes fully ionized matter, which is not suitable for planets. We compared this conductivity with the fit to the ab initio calculations for the electron conductivity of partially ionized water \citep[eq. 30 in][]{FrenchRedmer2017}. Their fit is valid in the same temperature density region as the work in \citet{French2019}.

Because MESA uses the opacity formulation of the thermal transport, we combined the contributions from vibrations and electrons as a sum of the conductivities and converted the result into an opacity. This opacity was then combined with the radiative opacity using the reciprocal sum to get the total opacity. 
To avoid truncation errors in the reciprocal sum due to very low conductivities, we imposed a lower bound of $k\geq10^{-6}$ W/m/K on all the conductivity values.
This threshold has little  effect on the planetary evolution since it is at least five orders of magnitude smaller than the lowest conductivity obtained in any of our models. 
To verify this, we performed a simulation with a reduced lower limit of $k\geq10^{-8}$ W/m/K and found a deviation of only $\sim10^{-9}$ R$_\oplus$ in radius after 10 Gyr.

\subsection{Model setups}\label{sec:ModelSetup}
Our models correspond to planets shortly after their formation process has been completed. 
The initial model was constructed given a mass, composition profile, and primordial entropy. We simulated the evolution for three different planetary masses of $M\low{p}=5\,,\, 10\,,$ and $15$ M$_\oplus$. To account for the uncertainty in the primordial planetary entropy, we considered three different central specific entropies of $s\low{center, i}=0.5$, $0.6$, and $0.7$ k$\low{B}/$m$\low{H}$, which we refer to as "cold," "warm," and "hot," respectively. 

\begin{figure}[hbt!]
    \centering
    \includegraphics[width=1\linewidth]{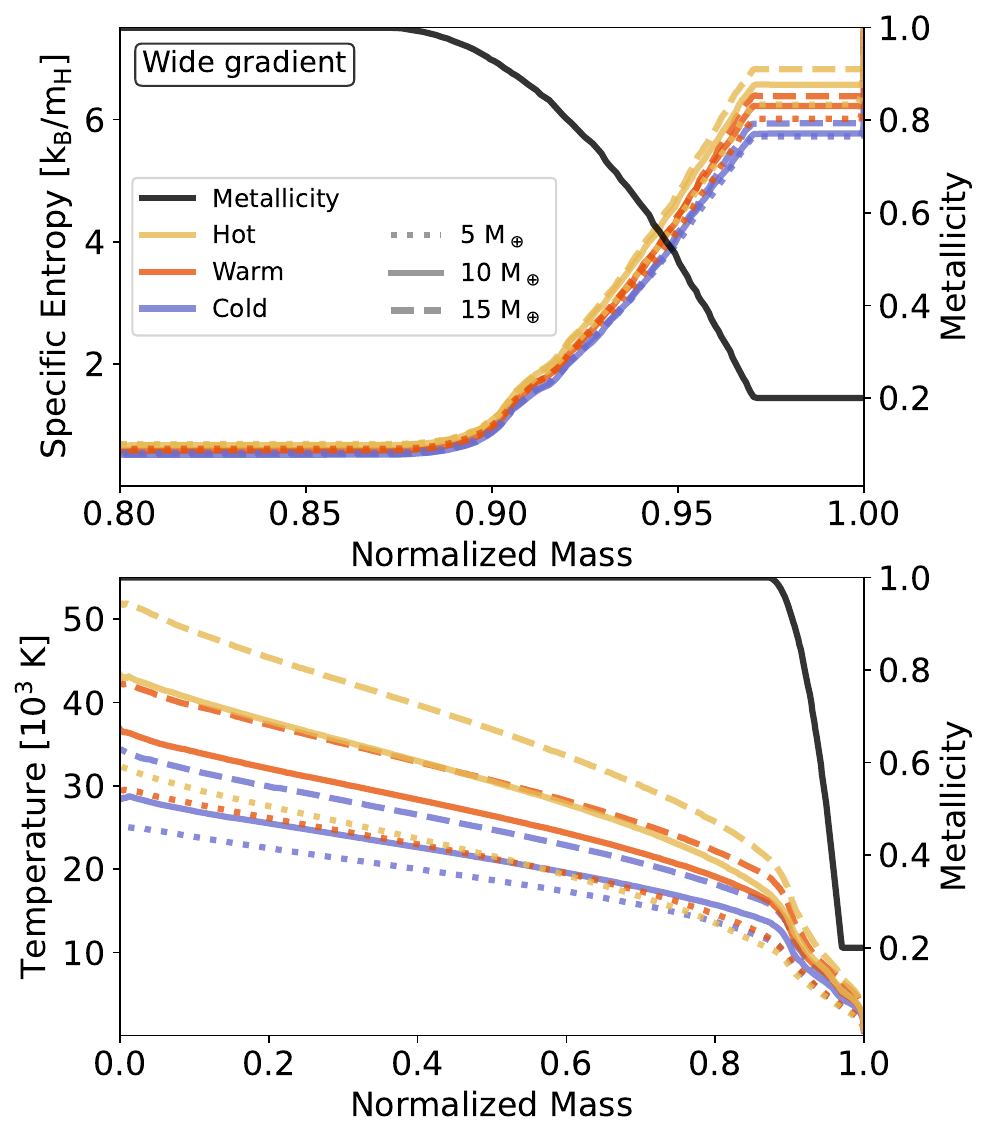}
    \caption{
    {Initial profiles for the wide composition gradient, showing specific entropy, temperature, and composition as functions of normalized mass. The colors blue, orange, and yellow correspond to different primordial entropies. The dotted, solid, and dashed lines correspond to planets with a mass of 5 M$_\oplus$, 10 M$_\oplus$, and 15 M$_\oplus$, respectively.}
    {\bf Top:} Specific entropy vs.~normalized mass of the initial model for the wide composition profile.
    The metallicity of the composition profile is shown by a black line, with its y-axis on the right-hand side. {\bf Bottom:} Temperature vs.~normalized mass of the initial model for the wide composition profile. The black line corresponds to the metallicity.}
    \label{fig:InitialEntropyCompostionAndTemperatureProfiles}
\end{figure}

\begin{table}[hbt]

    \caption{Models used for the conductivity by vibration and electrons. All models use the AESOPUS2.1 radiative opacity.}
    \begin{tabular}{cccc}
        \hline  
        Conductivity & $k\low{vib}$ & $k\low{elec}$ & References\\\hline
        Cond-1 & H$_2$O  & partially ionized H$_2$O & (1), (2)\\
        Cond-2 & - & fully ionized & (3)\\
        Cond-3 & - & constant $k\low{elec}=4$ $\frac{\text{W}}{\text{mK}}$ & -\\
        Cond-4 & MgSiO$_3$ & partially ionized H$_2$O & (1), (4)\\
    \hline    
    \end{tabular}
    \vspace{\baselineskip}
    
    References: 
    (1) \citet{FrenchRedmer2017},
    (2) \citet{French2019},
    (3)~\citet{Cassisi+2007},
    (4) \citet{Stamenkovic+2011}

    \label{tab:ModelSetups}
\end{table}
We created models with a wide and narrow composition gradient. Except for what we report in Section \ref{sec:ResultsNarrowCompGrad} and if not otherwise stated, we used the wide transition profile. The envelope metallicity was fixed to $Z\low{env}=0.20$ and the deep interior has $Z\low{int}=1.0$. The transition region has a width of $\Delta q=0.1$ in terms of the normalized mass coordinate $q$ for the wide gradient and $\Delta q=0.001$\footnote{The published version contained a typo, stating the width of the narrow gradient as $\Delta q =0.01$. The correct value used in the simulations is $\Delta q=0.001$.} for the narrow gradient. The position was fixed to achieve a bulk composition of $Z_\text{bulk}=0.95$ with a hydrogen-helium ratio of $Y/X=0.34$ based on the solar photosphere \citep{Asplund+2009}. 
Figure \ref{fig:InitialEntropyCompostionAndTemperatureProfiles}  shows the initial specific entropy, composition, and temperature profile for the wide composition gradient (see Figure \ref{fig:InitialEntropyCompostionAndTemperatureProfiles_dq0.01} for the narrow composition gradient). Details on how we used MESA's relax options to create our initial model are presented in Appendix \ref{sec:InitialModelMESA}. 

To compare the different approaches to the conductivity, we created multiple model cases: Cond-1 uses the vibrational and electronic conductivity for water under high pressure \citep{FrenchRedmer2017, French2019}, Cond-2 uses the default electron conductivity of MESA for fully ionized material based on \citet{Cassisi+2007}, Cond-3 assumes a constant electron plus vibration conductivity based on the conductivity at the Earth's core-mantle boundary of $k\low{elec}=4$ W/m/K \citep[e.g.][]{Stevenson+1983, Lobanov+2021}, and Cond-4 uses the phonon model for the vibrational conductivity \citep{Stamenkovic+2011} and the electron conductivity for partially ionized water \citep{FrenchRedmer2017}. All the cases use the AESOUPS tables for the radiative conductivity. We summarize the models in Table \ref{tab:ModelSetups}.
Overall, we recommend using Cond-1 because it is the most updated model. However, the conductivity of Cond-1 is currently only available for H$_2$O, and we hope that future work investigates the thermal conductivity of various materials under a large range of pressures and temperatures.

\section{Results}\label{sec:Results}
\subsection{The importance of thermal conductivity}\label{sec:CaseComparison}

\begin{figure}
    \centering
    \includegraphics[width=1\linewidth]{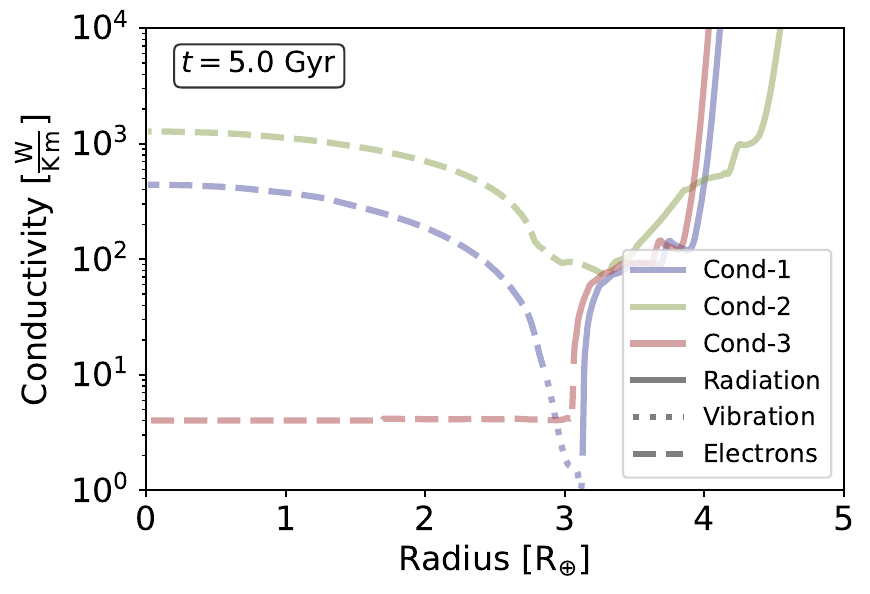}
\caption{Total conductivity vs.~ radius at $t=5$ Gyr for the warm $M_\text{p}$=10 M$_\oplus$ planet. The blue, green, and red curves represent Cond-1, Cond-2, and Cond-3, respectively. The largest contribution to the conductivity is shown by the line styles, i.e., solid, dotted, and dashed, for radiation (opacity), vibration, and electron dominated, respectively.}
    \label{fig:ConductivityRadiusProfilesM10Warm}
\end{figure}
We first compared the different conductivity cases: Cond-1, Cond-2, and Cond-3. We compared Cond-4 with Cond-1 separately because they only differ in a small region within the planet. We ran the evolution for 10 Gyr and compared the three cases. Figure \ref{fig:ConductivityRadiusProfilesM10Warm} shows the conductivity as a function of the planetary radius at $t=5$ Gyr. We chose this age to better reflect the typical ages of observed exoplanets, in contrast to the final simulation time of $t=10$ Gyr. The qualitative differences between the conductivities remained similar throughout the evolution. 

In the outer low-density parts of the planet, the radiative conductivity dominates. In the inner parts, where material is at least partially ionized, most of the thermal conduction is carried by electrons. In the intermediate part, the temperature is not high enough for the material to be ionized, but the material is too dense for photons to carry the energy. In this part, the vibration of nuclei and molecules take over. Although it is the most effective means of non-convective energy transport in this temperature-density regime, it is still inefficient. We note that only Cond-1 includes a model for the vibrational part, but assuming a fully ionized electron conductivity, as we did in Cond-2, makes the vibrational part insignificant. In Appendix \ref{sec:Cond1Uncertainty} we show how the uncertainty in Cond-1 affects the inferred radius of a $M_\text{p}$=10 M$_\oplus$ planet. 

In Figure \ref{fig:ConductivityRadiusProfilesM10Warm}, the difference in the outer part of the planet (from $\sim$ 3 R$_\oplus$) is caused by the different temperature-density regime of the radiative-opacity tables.
 The inner part, which is mostly dominated by the electron conductivity, is similar in shape for Cond-1 and Cond-2 but different by roughly one order of magnitude. In Cond-1, the planet has a region where the vibration of the molecules conducts most of the thermal energy. Yet, this mode is inefficient and leads to low conductivity.

\begin{figure*}[hbt]
    \centering
    \includegraphics[width=\linewidth]{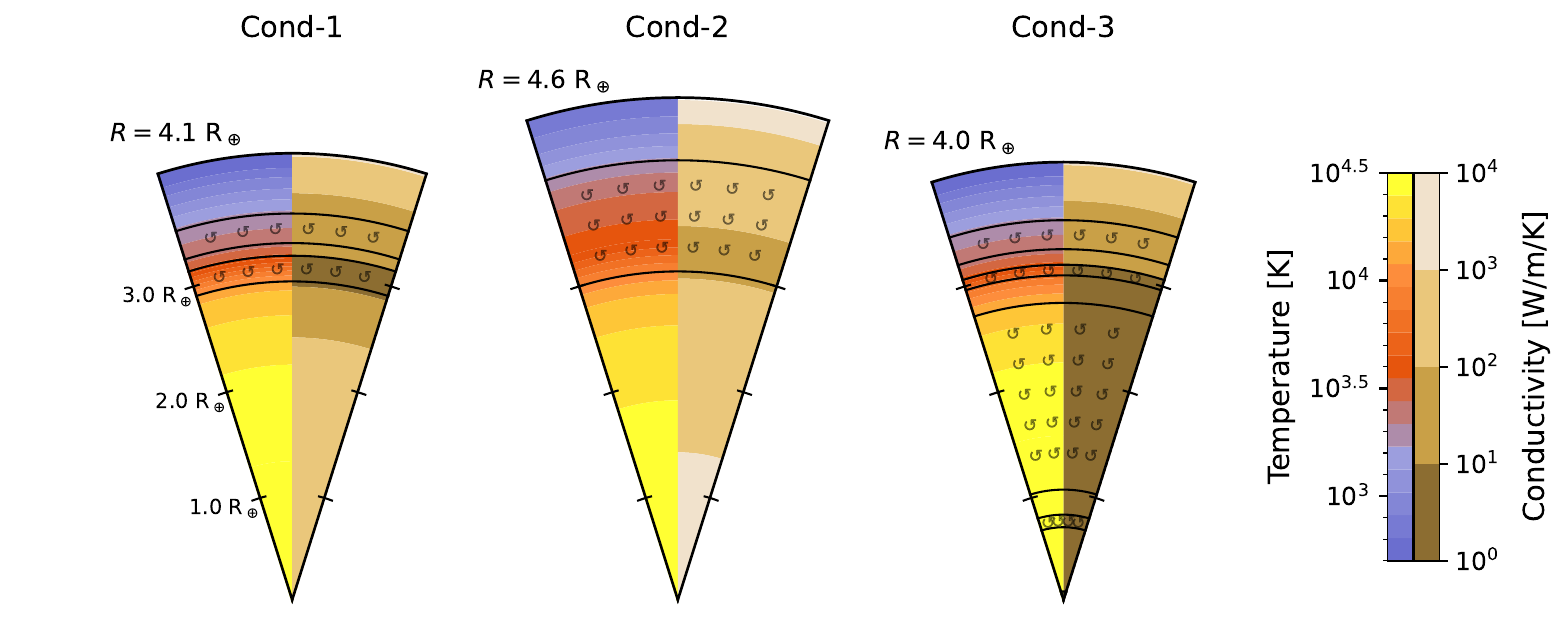}
    \caption{Temperature and conductivity as a function of the radius at $t=5$ Gyr for our three cases
    of the "warm" planet with $M_\text{p}=10$ M$_\oplus$. In each panel, the temperature is shown in the left half of the cone, colored in blue, red, and yellow from cold to hot. The order of magnitude of the conductivity is shown in the right half of each cone, colored in different shades of brown. The areas with circular arrows indicate convective regions within the planet. The 1 bar radius is also indicated.}
    \label{fig:FinalTemperaturComparison}
\end{figure*}

Next, we analyze the temperature profiles at $t=$ 5 Gyr in Figure \ref{fig:FinalTemperaturComparison} to explain the general energy transport and its effect on the planetary radius. In this study, the planetary radius is set to be the radius at the 1 bar pressure level. We also present the conductivity values throughout the interiors. The plots are intended to qualitatively illustrate which parts of the planet cool efficiently and which parts can contract or remain inflated. 
In Cond-2 the conductivity remains high throughout the planet.
Therefore, energy can be transported out of the deep interior efficiently. The large convective zone carries this energy toward the outer envelope and heats it from beneath. While the interior cools, the cooling of the outer envelope is controlled by the efficiency of the radiative cooling through the atmosphere, resulting in a generally hotter outer envelope. Because the outer envelope consists mostly of H-He, which have a high thermal expansion coefficient, the higher temperature envelope remains puffy throughout the entire evolution.  
On the other hand, in Cond-1 and Cond-3 the energy is efficiently trapped beneath non-convective regions with low conductivity. Both cases have hot interiors with temperatures above $10^{4.4}$K beyond a radius of $r\sim2$R$_\oplus$, which is a much larger region compared to Cond-2. 
Only a small amount of energy reaches the outer envelope, and thus the outer envelope can cool down efficiently and contract, resulting in smaller radii for Cond-1 and Cond-3 in comparison to Cond-2. 

\begin{figure}
    \centering
    \includegraphics[width=0.9\linewidth]{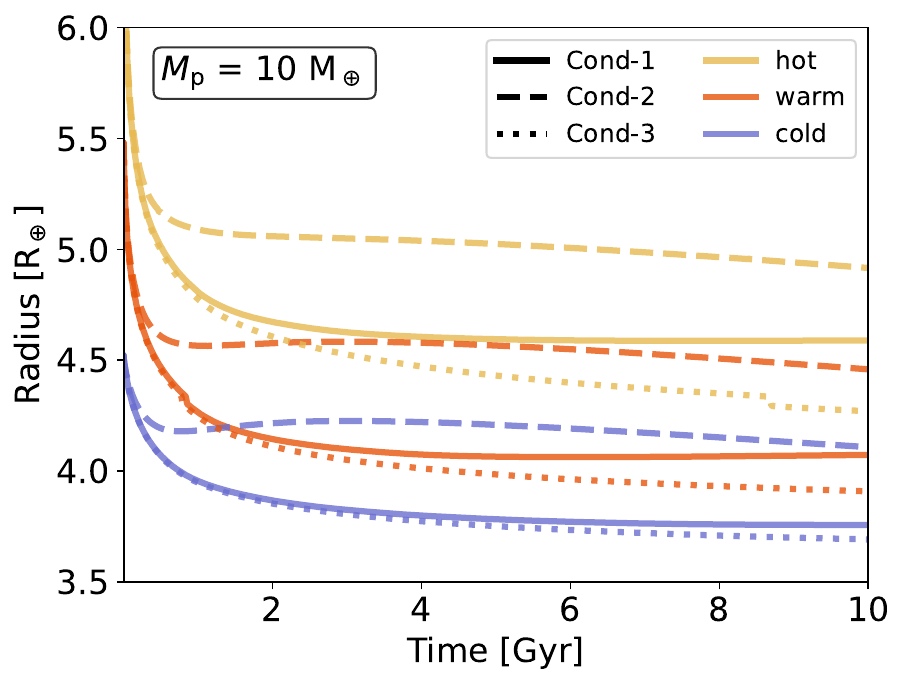}
    \caption{Radius vs.~time for a planet with $M\low{p}=10$ M$_\oplus$. The solid, dashed, and dotted lines correspond to Cond-1, Cond-2, and Cond-3, respectively. The different colors correspond to the different initial energy states.}
    \label{fig:RadiusEvolutionM10}
\end{figure}

Next, we study the evolution of the planets.
Figure \ref{fig:RadiusEvolutionM10} shows the radius against time for three initial energy states and a planet mass of $M\low{p}=10$ M$_\oplus$ (see Figure \ref{fig:RadiusEvolutionM05AndM15} in Appendix \ref{sec:MoreRadiusEvolution} for the $5$ M$_\oplus$ and $15$ M$_\oplus$ mass planets). We find the same phases in the evolution of a planet with a thermal boundary layer as \citet{Scheibe+2021}. 
The envelope first cools down and contracts, depending on the flux emitted throughout the atmosphere. This phase occurs on a timescale of 0.1 - 1 Gyr.
Next, the evolution is governed by the flux from the deep interior to the outer envelope throughout conductive layers. The balance between the energy that is released throughout the atmosphere and the energy that is received from the interior determines the size of the planet (i.e., how inflated it is). Then the contraction is mainly determined by the contraction of the deep interior.  However, the high-$Z$ material has a low thermal expansion coefficient such that cooling does not lead to a large contraction and the planet evolves slowly.
In the last phase, the energy reservoir of the deep interior is depleted, and contraction accelerates again.

We highlight that Cond-1 and Cond-3 are very similar during their early evolution and deviate more as time progresses. Initially, the contraction is very fast until it slows down after $t\approx2$ Gyr. 
The initial contraction is caused by the cooling of the outer envelope and stalls as soon as the internal heat prevents further cooling of the envelope.  
The planetary radius then depends on the heat flux from the deep interior to the outer envelope. Cond-2 is very different from the other two cases and has a larger radius during most of the evolution. Apart from the outer region of the planet, the conductivity in Cond-2 is higher than in the other cases. The higher conductivity leaks more energy from the deep interior to the outer envelope. This keeps the envelope hotter and therefore more extended compared to the cases with lower conductivities. As a result, the envelope is  inflated throughout the entire planetary evolution in this case.

\begin{figure*}[hbt]
    \centering
    \includegraphics[width=\linewidth]{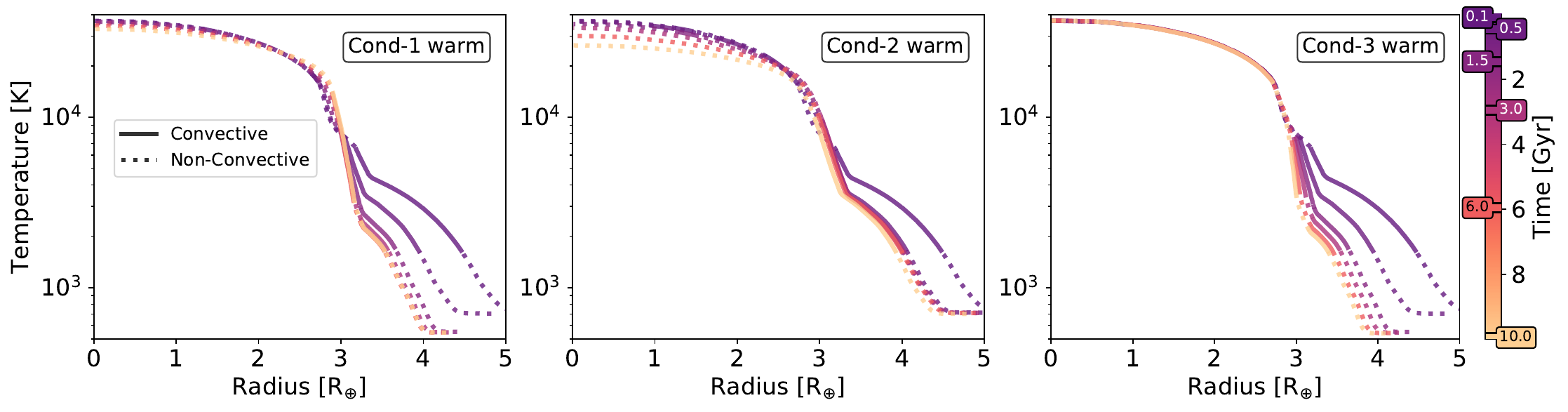}
    \caption{Planetary temperature profile at various times. The three different conductivity cases are shown in the panels from left to right for the warm $M_\text{p}=10$ M$_\oplus$ model. The color indicates the planetary age, while the style of the line indicates whether the zone is convective (solid) or non-convective (dotted).}
    \label{fig:TemperatureRadiusProfilesM10Warm}
\end{figure*}

In order to understand which regions can cool and contract and which remain inflated, we show the temperature profiles at various times in Figure \ref{fig:TemperatureRadiusProfilesM10Warm}. 
In Cond-1, the temperature in the deep interior near the center decreases over time. Due to the slow energy transport into the envelope, the envelope contracts and increases the pressure on the layers beneath. This increasing pressure and the low thermal conductivity through the region with the composition gradient cause a slight increase in temperature around the region of $r=3$ R$_\oplus$.
The higher conductivity of Cond-2 leads to faster cooling of the deep interior. The profiles of the outer envelope change little after $t=0.5$ Gyr. The heat from the interior keeps the envelope more extended earlier than in the other cases.

In Cond-3, the planet cools even slower than Cond-1 in the deep interior. 
Since the heat remains trapped in the deep interior, the outer envelope can cool down and contract most efficiently in Cond-3.
As soon as the energy of the deep interior is depleted in Cond-1 and Cond-2, the planets can contract more, while Cond-3 still fuels its slightly heated envelope. 
Hence, we expect that if the simulation time were much longer, Cond-3 would lead to an inflated radius over a longer period of time than Cond-1 and Cond-2.

\subsection{A narrow composition transition}\label{sec:ResultsNarrowCompGrad}

\begin{figure}
    \centering
    \includegraphics[width=0.8\linewidth]{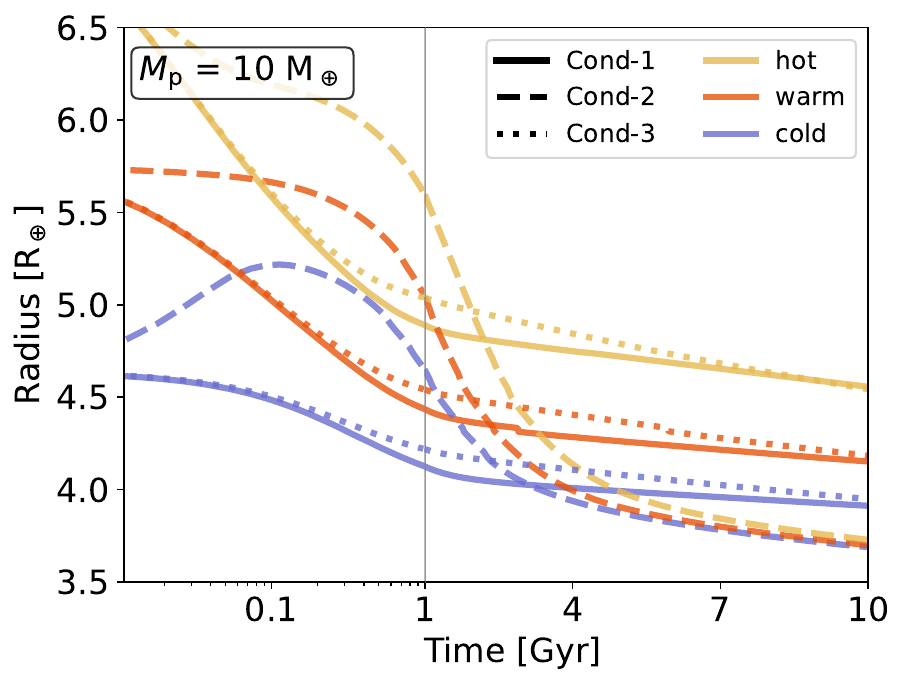}
    \caption{Planetary radius vs.~time for a planet with $M_\text{p}=10$ M$_\oplus$. The model is  similar to the one presented in Figure \ref{fig:RadiusEvolutionM10} but has a much  narrower composition transition ($\Delta q=0.001$). The solid, dashed, and dotted lines correspond to Cond-1, Cond-2, and Cond-3, respectively. The different colors show the different initial energy states. We note that the x-axis is logarithmic between 0.01 and 1 Gyr and linear between 1 and 10 Gyr.}
    \label{fig:RadiusEvolutionM10dq0.01}
\end{figure}

Next, we investigate the effect of the conductivity value for a model with a narrow composition transition between the envelope and the interior metallicity. The structure is similar to a differentiated interior with an envelope of H-He and heavy elements on top of a H-He-free interior. Figure \ref{fig:RadiusEvolutionM10dq0.01} shows the radius evolution for the three conductivity cases for the warm 10 M$_\oplus$ planet. As the evolution occurs faster than for the wide composition transition models, we show a logarithmic time axis for $t\leq1$ Gyr to emphasize the behavior at early times. The narrow composition region leads to a small region that is dominated by thermal transport in a non-convective region, and therefore, energy is transported much more efficiently. 

Compared to the wide composition transition models we considered, Cond-1 and Cond-3 cool faster, and contraction continues throughout the entire evolution simulation. Cond-3 is slightly enlarged compared to Cond-1 for most of the evolution, and they converge to similar radii at 10 Gyr.  

The high conductivity of Cond-2 allows for very efficient cooling.  
The initial cooling phase, where only the envelope contracts, occurs on a timescale of 0.1 Gyr. Before $t=1$ Gyr, most of the internal energy is depleted such that the flux into the envelope decreases and the planet can contract again. Interestingly, the cold Cond-2 model (dashed blue) expands significantly due to the energy flux from the deep interior. 
We find that all the initial energy states for Cond-2 converge to a similar radius within 10 Gyr. After that, the contraction is governed by the radiative cooling of the atmosphere.

\begin{figure*}[hbt]
    \centering
    \includegraphics[width=\linewidth]{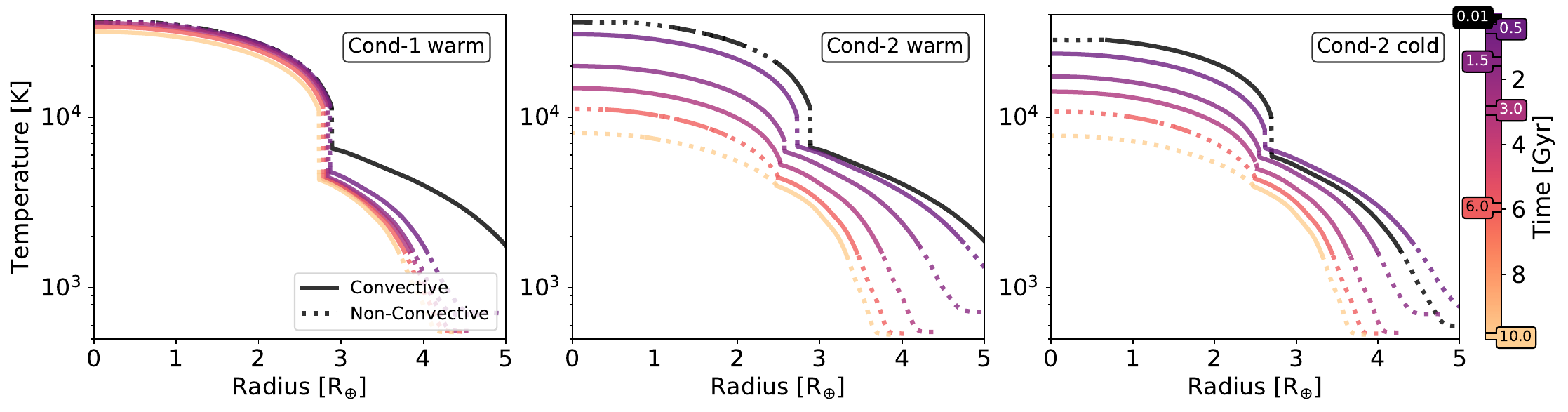}
    \caption{Temperature vs.~radius at different times for the narrow composition transition. We show the warm Cond-1 and Cond-2 in the left and middle panel and the cold Cond-2 in the right panel for the $M=10$ M$_\oplus$ model. The color indicates the  planetary age, and the solid (dashed) line indicates whether the zone is convective (non-convective).}
    \label{fig:TemperatureRadiusProfilesM10dq0.01}
\end{figure*}

To better demonstrate the cooling behavior of the interior, Figure \ref{fig:TemperatureRadiusProfilesM10dq0.01} shows the warm Cond-1 and Cond-2, and the cold Cond-2 temperature-radius profile for various times. Since Cond-3 is similar to Cond-1, it is not shown.  

In the case of Cond-1, the sharp composition change leads to a clear layered thermal profile. 
Two large convective layers are separated by a non-convective layer that is stabilized by the composition gradient. 
The outer envelope is again non-convective because of efficient energy transport by radiation. After the initial contraction of the outer envelope, the slow-cooling deep interior acts as an energy reservoir that keeps the envelope inflated for several gigayears.

In Cond-2, the thermal energy transport is sufficiently high to decrease the internal energy reservoir by an order of magnitude. After most of the energy is emitted, the outer envelope can cool down and contract much more than in Cond-1 and Cond-3. As mentioned above, the radius of the cold Cond-2 model even increases in the beginning. The early profile at $t=0.01$ Gyr (black lines in Figure \ref{fig:TemperatureRadiusProfilesM10dq0.01}) has a more contracted envelope than the slightly more evolved profile at $t=0.5$ Gyr. The central temperature drops by almost $5,000$ K within these two time steps, releasing more energy into the outer parts than the atmosphere can emit. At $t\sim5$ Gyr the temperature gradient across the composition transition approaches zero. The deep interior becomes non-convective, as the temperature gradient in the deep interior is lower than the adiabatic gradient. At this point, the evolution is governed by the thermal flux that escapes via the planetary atmosphere.

\subsection{Vibrational conductivity model comparison}
\begin{figure}[hbt]
    \centering
    \includegraphics[width=1\linewidth]{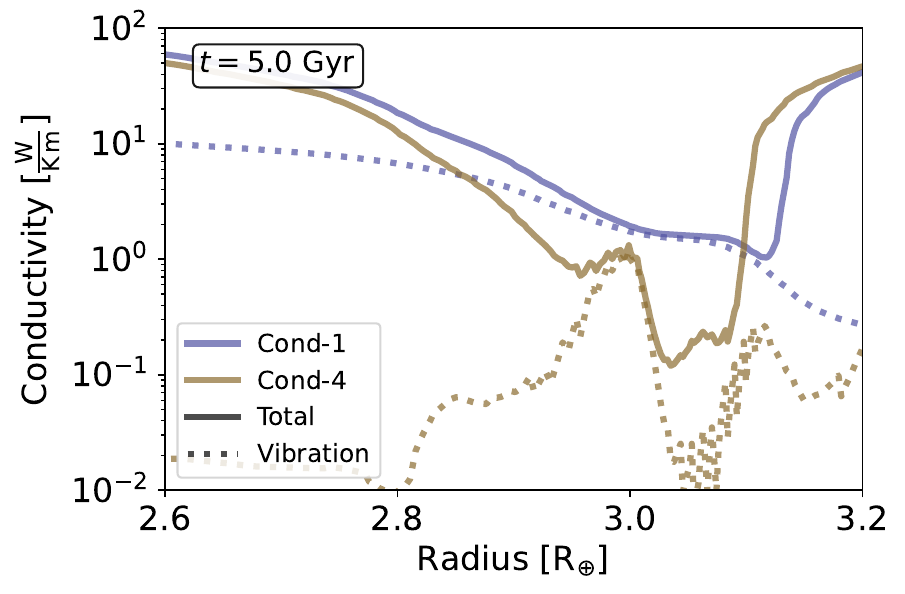}
    \caption{Conductivity vs.~radius at $t=5$ Gyr for the warm $M\low{p}=10$ M$_\oplus$ planet. The blue and brown lines correspond to Cond-1 and Cond-4, respectively. The total conductivity is shown by the solid line, and the thick line shows the contribution from the vibrational conductivity. Outside the shown radius window, the conductivity is dominated by the electron and radiation contribution.}
    \label{fig:ConductivityRadiusProfilesM10WarmPhonon}
\end{figure}

Next, we compare the models Cond-1 and Cond-4, which differ only in the treatment of the vibrational conductivity. 
Figure \ref{fig:ConductivityRadiusProfilesM10WarmPhonon}  shows the total conductivity and the vibrational component at $t=5$ Gyr for the warm $M_\text{p}=10$~M$_\oplus$ planet.
In Figure \ref{fig:RadiusEvolutionPhononsM10} we show the radius over time for Cond-1 and Cond-4. 
We find only small variations in all the tested masses and initial energy states, where the largest difference is found to be $1.6\%$ for the hot $M_\text{p}=10$~M$_\oplus$ model. The key difference lies in the smaller conductivity in Cond-4.
We note that we used the same electronic conductivity in Cond-1 and Cond-4. A different model for the electronic conductivity can increase the region in the planet where the vibrational conductivity dominates and therefore increase the deviations between Cond-1 and Cond-4.
We also note that the scaling law assumed in Cond-4 might not be suitable for the density range in the planet, where the vibrational conductivity dominates. This is because in this region, the densities are lower by about a factor of ten in comparison to the reference conductivity value of $\rho\low{ref}=3.87$ g cm$^{-3}$. More information on the vibrational conductivity at various conditions is therefore important.

\subsection{Implications for exoplanet characterization}
\begin{figure}[hbt]
    \centering
    \includegraphics[width=0.9\linewidth]{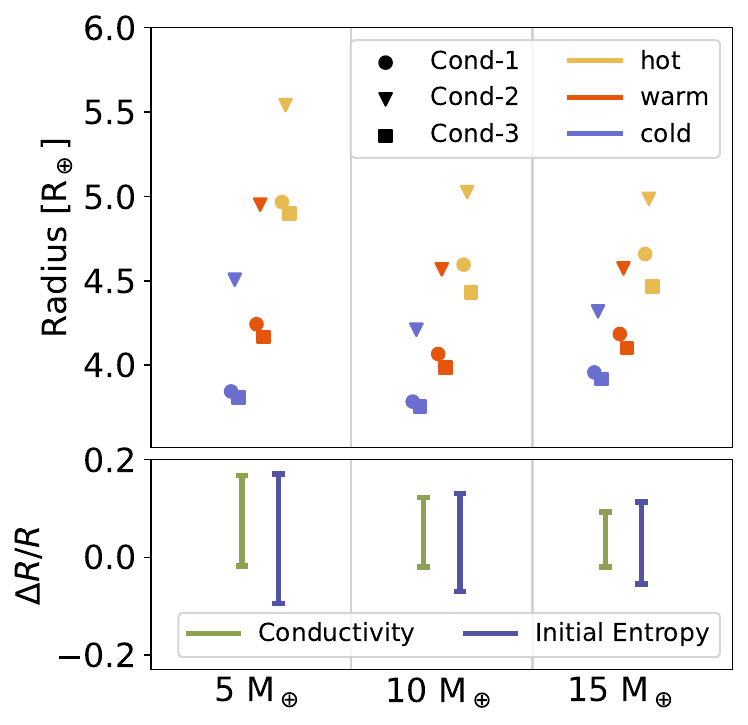}
    \caption{
    {Planetary radius at $t=5$ Gyr and relative difference between the models and initial entropies.}
    {\bf Top:} Planetary radius at $t=5$ Gyr for the models with the wide composition gradient. Different markers indicate the different conductivity cases, while the different colors correspond to the different initial entropies. The three planet masses ($M\low{p}=5$, 10, and 15 M$_\oplus$) are shown from left to right.
    {\bf Bottom:} Relative difference of the inferred radius $\Delta R/R = (R' - R)/R$. Green corresponds to the  deviation caused by the different assumed conductivities. We compare the warm Cond-1 with the warm Cond-2 and Cond-3. Blue corresponds to the  deviation caused by different initial entropy states. We compare the warm Cond-1 with the cold and hot Cond-1.}
    \label{fig:RadiusAt5Gyr}
\end{figure}
We found large differences in the inferred radii depending on the model assumptions. This has important implications for the characterization of exoplanets in the relevant mass regime. Below, we analyze the results for the models with the wide composition gradient. Figure \ref{fig:RadiusAt5Gyr} shows the planetary radii at $t=5$ Gyr for the first three conductivity cases; the cold, warm, and hot initial entropy states; and the three masses (see Table \ref{tab:RadiiAt5Gyr} for the exact values of the radii).
 The bottom panel shows the relative difference in the inferred radius when considering  different conductivity models and initial entropies. In particular, we compare the warm Cond-1 with the warm Cond-2 and warm Cond-3 models. For the comparison in the inferred primordial  entropy, we compare the warm Cond-1 with the cold and hot Cond-1 models. 
We note that all models have the same composition profile and the same bulk composition, yet the evolution still leads to very different radii. Therefore, for a given mass, the radius uncertainty is found to be of the order of $\sim20\%$, depending on the conductivity, and of the order of $\sim25\%$, depending on the primordial entropy, when considering the spread between the minimum and maximum values in Figure \ref{fig:RadiusAt5Gyr}. 
These differences are much larger than the typical measured uncertainty of the planetary radius, which is of the order of $\sim 5\%$ and could go even lower with upcoming missions (e.g., PLATO). 
As a result, it is clear that theory plays a key role in the characterization of Neptunes and sub-Neptunes and that in order to take full advantage of the data, improvements in theory are required. In particular, making such improvements is critical to determining the conductivities of various elements at the relevant pressure-temperature regime and to better constraining the post-formation entropy (and composition gradient) of such planets. 
We hope that future studies will focus on these topics. 

\begin{table}[hbt]
    \caption{Inferred radius in units of Earth's radii at $t=5$ Gyr for the models with the wide composition gradient.}
    \centering
    \begin{tabular}{cc|ccc}
        \hline
        {\bf Mass} & {\bf Conductivity} 
        & {\bf Cold} & {\bf Warm} & {\bf Hot} \\\hline
                           \text{ }&           \\
        \multirow{3}{*}{5 M$_\oplus$} & Cond-1 & 3.8 & 4.2 & 5.0\\
        
                                      & Cond-2 & 4.5 & 5.0 & 5.5\\
                                      & Cond-3 & 3.8 & 4.2 & 4.9\\
                           \text{ }&           \\
                           \hline
                           \text{ }& \\
                           
        \multirow{3}{*}{10 M $_\oplus$} & Cond-1 & 3.8 & 4.1 & 4.6\\
                                      & Cond-2 & 4.2 & 4.6 & 5.0\\
                                      & Cond-3 & 3.8 & 4.0 & 4.4\\
                         \text{ }&               \\
                         \hline
                         
                       \text{ }& \\
        \multirow{3}{*}{15 M$_\oplus$} & Cond-1 & 4.0 & 4.2 & 4.7\\
                                      & Cond-2 & 4.3 & 4.6 & 5.0\\
                                      & Cond-3 & 3.9 & 4.1 & 4.5
 
    \end{tabular}
    \label{tab:RadiiAt5Gyr}
\end{table}

\section{Discussion}\label{sec:Discussion}
While our study presents a step forward toward  the understanding of Neptunes and sub-Neptunes, some simplifying assumptions have to be made, which we discuss below. 
First, we note that the planetary cooling strongly depends on the atmospheric model. In this work we used a standard semi-gray irradiated atmosphere model \citep{GuillotHavel2011}. 
Our choice of atmospheric model represents a compromise between accuracy and complexity. The semi-gray approximation has been shown to deviate by $\lesssim10\%$ from full numerical solutions to the radiative transfer problem across a wide range of gravities and effective temperatures \citep{Parmentier+2015}.
However, a different model would change the inferred thermal energy flux through the atmosphere, leading to a different temperature at the bottom of the atmosphere. If this flux is significantly different compared to the one we calculate here, the radius evolution would also be affected.
Since the contraction of the outer envelope depends on the energy balance between the flux coming from the deep interior and the flux that is emitted through the atmosphere, our results are still valid in terms of the relative differences between different model assumptions. 

Second, we have presented low-temperature opacity tables using the AESOPUS2.1 code \citep{Marigo+2024}. 
The new tables cover a range of $2\leq\log(T/$K$)\leq4.5$ for different metallicities and a constant $Y/X=0.34$ fraction. They are an improvement of the commonly used Freedman opacities \citep{Freedman+2008} included in MESA since they cover a higher temperature range. Their disadvantage compared to the Freedman opacities is the limited range in the parameter $\log R=\log\rho- 3\log T + 18$ (all units in cgs). The Freedman tables cover a range of up to $\log R \leq 9$, while the AESOPUS2.1 code only supports values up to $\log R \leq 6$. The planet models in this study can reach values of $\log R \sim 7 - 8$, and we hope to further improve on this in future work.  In cases where $\log R$ is out of the range of the tables, MESA uses the last available grid point. This means that for a given temperature and density with a higher value of $\log R>6$,  the opacity is used for the given temperature and a lower density to match the last available $\log R=6$. 

Third, for simplicity, we kept the composition profile fixed over time and only considered the effect of the energy transport from convection.
In reality, convective mixing erodes composition gradients \citep[e.g.][]{KnierimHelled2024, TejadaArevalo2025}.
Including the effect of convective mixing would affect the inferred evolution, especially for the composition profile with a wide transition region. 
Such a composition gradient could mix from outside inward, and the composition may start building a step-like profile with narrow non-convective regions separating convective regions. 
As already stated in the introduction, Uranus is suspected to have a stable non-convective boundary layer not too far inside its interior. Furthermore, if a planet forms cold (low entropy), mixing is likely to be less efficient, and stable composition gradients can be expected in many planets.
Therefore, as long as there is a stable non-convective region, it is clear that the choice of the conductivity is significant and should be carefully  considered. We hope to further improve on this choice in a future study that self-consistently treats convection and mixing.

Fourth, due to a lack of data, the assumed material of the EoS is not always the same as the material in our conductivity models. We took four choices for the conductivity models that all assume a different material. This material is different from the water-rock mixture we assumed for the EoS and does not scale with the metallicity. This inconsistency is hard to improve, as the data for the conductivities are sparse. We hope that further investigations of the thermal conductivity of different materials will provide the required information, allowing for more consistent planetary models.

Fifth, although we have investigated multiple cases with different masses, the initial temperature profile in the models is nearly adiabatic.
However, formation models suggest that composition gradients form during the planetary formation process \citep{VallettaHelled2020, VallettaHelled2022}. Therefore, Neptunes and sub-Neptunes could have temperature profiles that differ from the adiabatic ones after their formation. 
The composition profile we used here is based on the idea of having a smooth transition between the central metallicity and an envelope metallicity similar to what is expected from Uranus interior models. Because mixing likely starts from the outside in, we assumed a flat composition profile for the envelope, which might have been an eroded composition gradient. We suggest that future studies include large ranges of bulk metallicities and internal structures. 

Sixth, the planetary evolution depends on the choice of the EoS. In Appendix \ref{sec:RadiusComparisonEoS}, we compare the radius evolution obtained with different H–He EoSs and find that simulations using the SCvH EoS \citep{Saumon+1995} yield radii up to $\lesssim 2\%$ larger after 10 Gyr compared to those with the CMS EoS. Variations in the EoS of heavy elements would also affect the results. For example, \citet{Howard+2025} compared two water EoSs for static adiabatic models for Uranus, which leads to a temperature difference of $\sim$1000K at a few hundred kbar. 
Given the large uncertainty in the planetary composition and the unknown ratio between different elements (e.g., ice and rock), the uncertainty of the EOS is expected to be small. At the same time, we note that a systematic study of the  uncertainties arising from both the EoS and the assumed materials is desirable. 

Seventh, for simplicity, we did not consider demixing and settling, which are processes that could take place and affect the evolution of sub-Neptunes and Neptunes. 
For example, the phase separation of water \citep{BailyStevenson2021, CanoAmoros+2024, Howard+2025} or carbon-bearing species \citep{He+2022, Cheng+2023, Militzer2024} could delay the planetary cooling and lead to radius inflation. We also note that progress in characterizing material properties of various mixtures in planetary conditions is required. We hope to address these topics in future research.

Finally, we note that for planets with a significant amount of water in the deep interior, we used the easy-to-use fits for the conductivity from \citet{FrenchRedmer2017} for the electronic contribution and those of \citet{French2019} for the vibrational contribution. 
\citet{Scheibe+2021} employed the same models for thermal conductivity and conclude that the presence of additional materials would enhance the conductivity relative to pure water. Thus, using the conductivity of pure water provides a lower bound. For example, the higher electrical conductivity of ammonia compared to water could lead to higher thermal conductivity \citep{Ravasio+2021}. It is therefore clear that in order to perform reliable planetary evolution simulations, one should use the appropriate conductivity data. We suggest that it is highly desirable to investigate the properties of different H–C–N–O mixtures at planetary conditions. 
Our work shows that, similar to the EoS, the conductivity of material plays an important role in planetary evolution and therefore also in planetary characterization.  
\par

\FloatBarrier
\section{Summary and conclusions}\label{sec:SummaryAndConclusion}

Our Galaxy contains many Neptunes and sub-Neptunes, but their characterization remains a challenge. 
Often, these planets are modeled assuming a differentiated structure and adiabatic interior. In this work, we have investigated how the planetary thermal evolution, and therefore also the inferred internal structure, is affected when the planets consist of non-convective layers. 

We find that the treatment of the stable layer, the thermal conductivity in particular, significantly affects the results.  
We have shown that the commonly adopted simplifications, such as electron conductivity based on fully ionized hydrogen or a constant conductivity profile, lead to very different radius predictions.
In addition, we find that the vibrational contribution to the conductivity must be included.  
In particular, the treatment of the conductivity is critical if a significant fraction of the planet's total energy is trapped beneath a non-convective layer. 
The initial energy state also has a key effect on the planetary radius and its evolution. For a given mass and composition, we find that the inferred radius can differ by up to $\sim20\%$ depending on the model for the conductivity and by up to$\sim25\%$ depending on the primordial energy state. Our results show that knowledge of the conductivity (of various mixtures) at planetary conditions and further constraints on the primordial state of Neptunes and sub-Neptunes are crucial for the characterization of this planetary type. 

\begin{acknowledgements}
We thank the referee and editor for their helpful comments. We also  thank Simon Müller and Henrik Knierim for many valuable discussions and technical  support. The authors  acknowledge support from SNSF under grant \texttt{\detokenize{200020_215634}}.\\
\textit{Software:} MESA \citep{Paxton+2011, Paxton+2013, Paxton+2015, Paxton+2018, Paxton+2019, Jermyn+2023}, AESOPUS2.1 \citep{MarigoAringer2009, Marigo+2024}, Jupyter Notebook \citep{Kluyver2016, Granger+2021}, MesaReader, NumPy \citep{Harris2020}, Matplotlib \citep{Hunter2007}, Astropy \citep{Astropy2013, Astropy2018, Astropy2022}

\end{acknowledgements}

%

\bibliographystyle{aa} 
\bibliography{bibliography}

\begin{appendix}




\onecolumn
\section{A low-temperature opacity table with AESOPUS2.1}\label{sec:AESOPUSTable}
\begin{figure*}[!ht]
    \centering
    \includegraphics[width=1\linewidth]{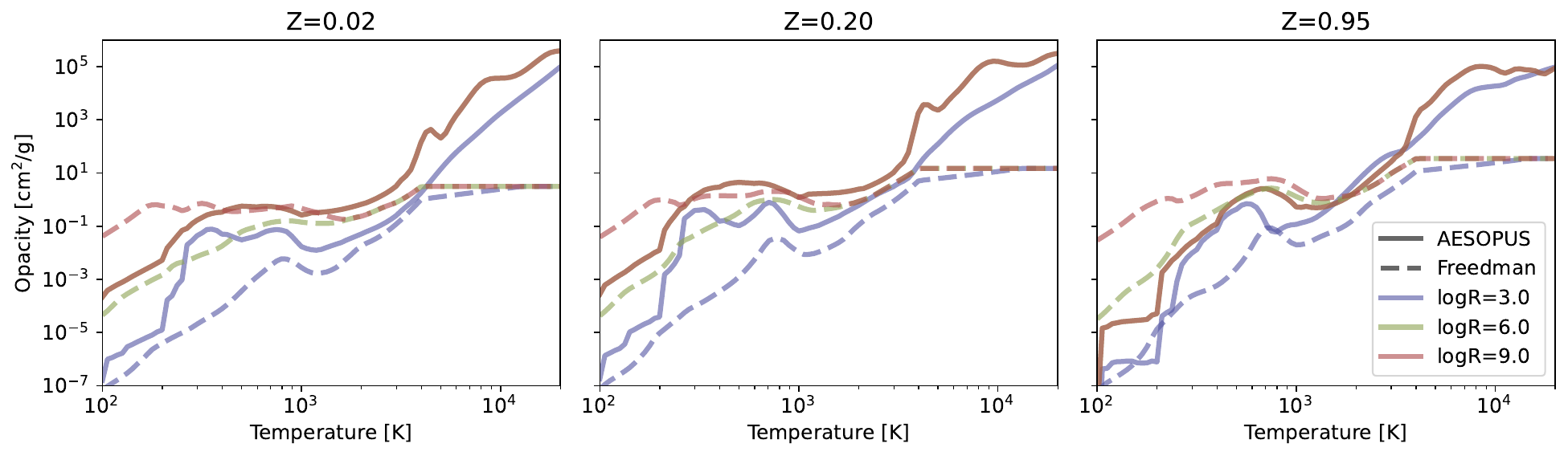}
    \caption{Radiative opacity as a function of temperature for constant $\log R$. The left, middle and right panels correspond to metallicities of $Z=0.02,\,0.20$, and 0.95, respectively. The dashed line shows the radiative opacity from \citet{Freedman+2014} as included in MESA and the solid line shows the custom-made AESOPUS2.1 tables using the web interface \citep{Marigo+2024}. The three different colors show different values for $\log R$. We note that the AESOPUS2.1 tables are only available up to $\log R=6$, resulting in the same AESOPUS2.1 opacities for $\log R\geq6$ for a given temperature (orange and green solid-line).}
    \label{fig:opacity_comparison}
\end{figure*}

We use the web interface of the AESOPUS2.1 code \citep{MarigoAringer2009, Marigo+2024} to create new low temperature radiative opacity tables that cover low and high metallicities. We use a fixed helium to hydrogen mass ratio of $Y/X=0.34$ corresponding to the solar photosphere \citep{Asplund+2009}. Following the Freedman tables included in MESA,  we create 7 tables for $Z\in\left\{0.01,0.02,0.04,0.1,0.2,0.63,0.95\right\}$.
Due to numerical limitations, we set the maximum heavy-element mass fraction to $Z=0.95$. For the elemental abundance ratios, we select the reference solar composition of \citet{Lodders2003}. 

We use the full available temperature range of $2\leq\log T \leq4.5$ in steps of $\Delta\log T=0.02$ between $2\leq\log T\leq3.5$ and $\Delta\log T=0.05$ between $3.5\leq\log T\leq4.5$. The density range is given by $-8 \leq\log R\leq6$ in steps of $\Delta\log R = 0.2$. 
The Web interface does not allow one to calculate the full grid in one query. Therefore, we split the temperature range into three regions and manually merge the tables. The tables provided by \citet{Freedman+2014} are available up to $\log R=9$, while the AESOPUS2.1 tables are limited to $\log R=6$ at a given temperature. When a requested $\log R$ falls outside the available range, MESA’s opacity module keeps the temperature fixed and substitutes a lower density so that the inferred opacity corresponds to the highest available $\log R$ value.  
Figure \ref{fig:opacity_comparison}  shows the opacity as a function of the temperature for constant $\log R$ and compare our new table with the Freedman opacity. 
\citet{Marigo+2024} suggested that the differences between their results and those of \citet{Freedman+2014} could be due to pressure broadening and a lower number of molecules considered for molecular transitions.

\section{Vibrational conductivity}\label{sec:VibrationalConductivity}
\citet{Ross+1984} reviewed the phonon model for a completely anharmonic crystal and derived the temperature and density scaling of the thermal conductivity of liquids under high pressure. We follow the formulation of \citet{Stamenkovic+2011}, such that the partial derivatives of the vibrational conductivity $k\low{vib}$ are given by 
\begin{align}
    \left(\parfrac{\ln k\low{vib}}{\ln T}\right)_\rho &= - 1,\\
    \psi:=\left(\parfrac{\ln k\low{vib}}{\ln\rho}\right)_T &= 3\gamma - 2 \left(\parfrac{\ln \gamma}{\ln\rho}\right)_T - \frac{1}{3},\label{eq:kvib_rho_scaling}
\end{align}
with $\gamma=\left(\partial\ln\nu/\partial\ln\rho\right)_T$ being the Grüneisen parameter for the average vibrational frequency $\nu$. The parameter $\gamma$ can be approximated using the thermodynamic Grüneisen parameter 
\begin{equation}
    \gamma\low{TH}=\frac{\alpha K_T}{\rho c_V},
\end{equation}
where $\alpha=-\left(\partial\ln\rho/\partial\ln T\right)_P$ is the thermal expansivity, $K_T=-\left(\partial\ln\rho/\partial\ln P\right)_T$ the isothermal compressibility, and $c_V$ the specific heat capacity at constant volume. Given a reference conductivity $k\low{vib, ref}(T\low{ref},\rho\low{ref})$ and an EOS to calculate $\gamma\low{TH}$, one can construct the vibrational conductivity from 
\begin{equation}
    k\low{vib} = k\low{vib, ref} \left(\frac{\rho}{\rho\low{ref}}\right)^\psi \left(\frac{T\low{ref}}{T}\right),
\end{equation}
with $\psi$ defined by eq. (\ref{eq:kvib_rho_scaling}) using the approximation $\gamma\approx\gamma\low{TH}$.

\clearpage
\twocolumn
\section{Initial model}\label{sec:InitialModelMESA}
In this section, we describe how we use MESA to produce the initial models. 
The basic outline is to first create an adiabatic model and utilize MESA's relax options to create the desired initial state. The procedure ensures that the initial model has a given mass $M\low{p, i}$, a specific composition profile $Z\low{i}(q)$ as a function of the normalized mass coordinate $q$, and an initial energy budget defined by some central specific entropy $s\low{central, i}$. If not stated otherwise, we let MESA run for 1000 years with maximal time steps of 100 years, between each step, to allow the model to settle in for better convergence in the long run.
\begin{enumerate}[wide, labelindent=0pt, itemsep=\baselineskip, label={\textit{Step \arabic*:}}, ref={\textit{Step \arabic*}}]
\item 
We create an adiabatic model using the option \verb|create_initial_model| with $M_p=20$ M$_\oplus$, $R_p=9.4R_\oplus$, $Z\low{bulk}=0.8$, $Y/X=0.34$. Note, that this model has a flat composition profile. We force the model to follow the adiabat given by the EoS.

\item \label{item:StepZCore}
Next, we relax the composition to $Z\low{bulk}=0.94$ using \verb|relax_initial_Z| to avoid convergence issues for planets with masses of $M\low{p}\lesssim 8$~M$_\oplus$. Ideally, one would set the metallicity in this step to the final central metallicity to ensure a consistent entropy, which is set in \ref{item:StepEntropy}. However, this led to issues, which is why we set $Z\low{bulk}$ to a lower value. This is accounted for by setting a slightly higher entropy in \ref{item:StepEntropy}. The model is forced to follow the adiabat.

\item 
Now we use the relax option \verb|relax_initial_mass| to shrink the current model down to its desired mass $M\low{p, i}$. This step can be difficult if $M\low{p, i}$ is low. In this case, increasing the metallicity as described in \ref{item:StepZCore} can help. Or before changing the mass one can set a different entropy profile as described in \ref{item:StepEntropy}. Again, we force the model to follow the adiabat.

\item \label{item:StepEntropy} 
We set the initial energy budget, which we parametrize by the central specific entropy $s\low{center, i}$. Using the option \verb|relax_initial_entropy| MESA applies an extra specific heat $Q_l$ in each layer $l$ according to 
\begin{equation}
    Q_l =  \frac{1 - s_l}{s\low{i}}\frac{E_l}{\tau\low{relax entropy}},
\end{equation}
with $s_l$ being the current specific entropy, $E_l$ the current specific energy and $\tau\low{relax entropy}=1\E{-3}$ yr an relax timescale. Usually, the adiabat, i.e., a constant entropy profile, agrees with the profile that is predicted by the mixing length theory. But especially for the outer parts the entropy profile can significantly deviate from an adiabat. This can lead to convergence issues where the solver reacts with rapidly changing profiles. We avoid this issue by forcing the model to follow the adiabatic gradient given by the EoS. The specific entropy depends on the metallicity, and our final composition profile features a central metallicity of $Z\low{int, i}=1$, which differs from the value set in step \ref{item:StepZCore}. As a result, in this step we assign a higher entropy such that the initial model after relaxing the composition has the desired entropy.

We construct three initial energy states — hot, warm, and cold — corresponding to central entropies of $s\low{center, i}=0.7\,,\, 0.6$ and $0.5$ $k\low{B}/m\low{H}$ respectively. Although models with different masses and narrow composition gradients may show slight variations in $s\low{center, i}$ they agree to within rounding to the first decimal place.

\item \label{item:StepCompGrad} 
Next, we set the composition profile using the option \verb|relax_initial_composition|. MESA relaxes the composition profile of the planet by reading in a file that contains the normalized mass coordinates for each layer with the desired mass fraction of each species. We use a reduced isotope network that only includes h1, h4, and o16 representing $X, Y$, and $Z$, respectively. This step sometimes leads to convergence issues. It is advised to keep the parameter \verb|num_steps_to_relax_composition| not too high and the relax timescale \verb|timescale_for_relax_composition| short. We find values with $50$ and $-1$ to be well-behaved.
Because the entropy is a function of temperature, density, and composition, the final profile will not have the flat entropy profile we set in \ref{item:StepEntropy}. 
 \begin{figure}[t]
    \centering
    \includegraphics[width=1\linewidth]{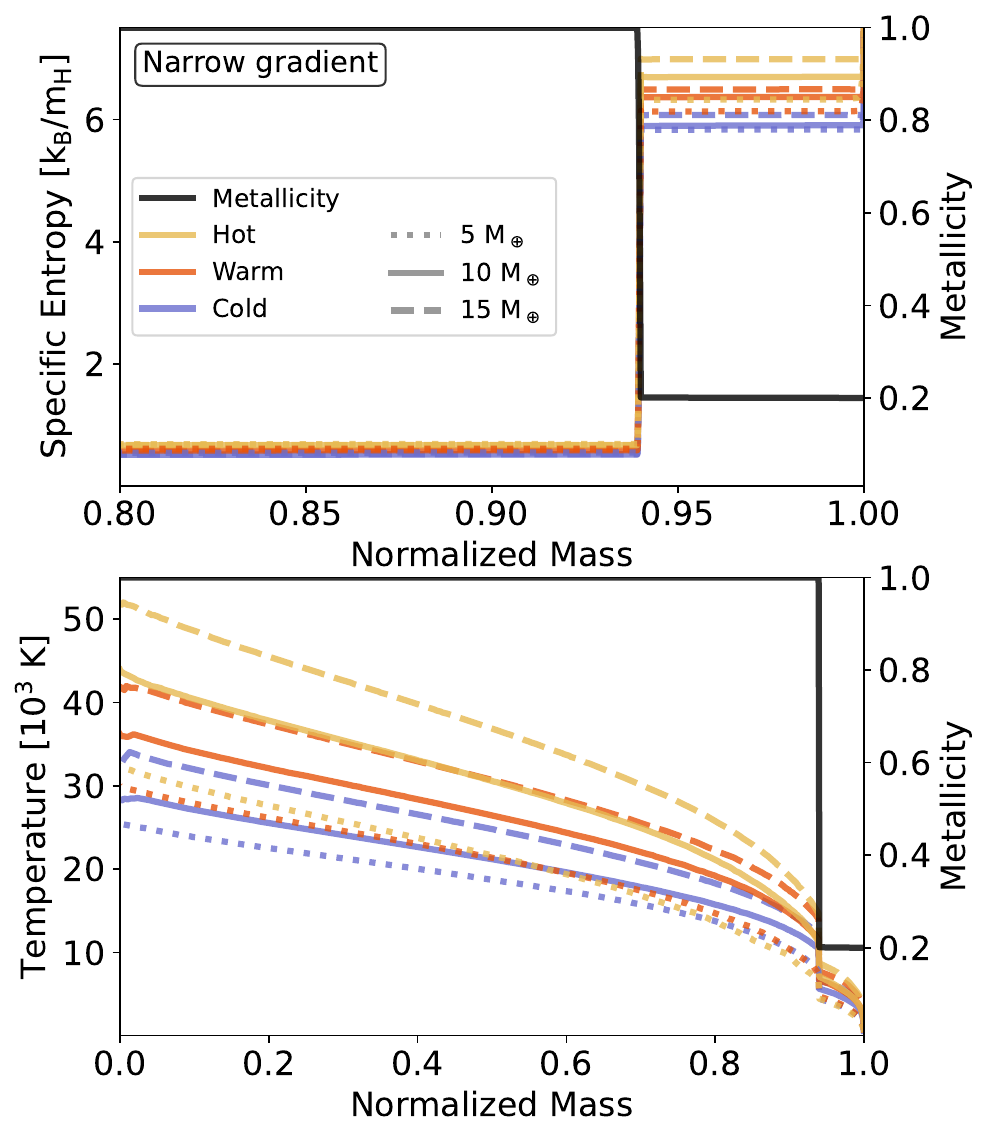}
    \caption{
    {Initial profiles for the narrow composition gradient, showing specific entropy, temperature, and composition as functions of normalized mass. The colors blue, orange, and yellow correspond to different primordial entropies. The dotted, solid, and dashed lines correspond to planets with a mass of 5 M$_\oplus$, 10 M$_\oplus$, and 15 M$_\oplus$, respectively.}
    {\bf Top:} Specific entropy vs.~normalized mass of the initial model for the narrow composition profile. The blue (cold), orange (warm), and yellow (hot) colors represent different initial entropies. The dotted, solid, and dashed lines correspond to planets with a mass of 5 M$_\oplus$, 10 M$_\oplus$, and 15 M$_\oplus$, respectively. The metallicity of the composition profile is represented by a black line, with its y-axis on the right-hand side. {\bf Bottom:} Temperature vs.~normalized mass of the initial model for the narrow composition profile. The black line shows the metallicity.}
    \label{fig:InitialEntropyCompostionAndTemperatureProfiles_dq0.01}
\end{figure}
The initial specific entropy profiles and composition profiles (after the completion of the four steps described above) are shown in the upper panels of Figure \ref{fig:InitialEntropyCompostionAndTemperatureProfiles} for the wide gradient. Figure \ref{fig:InitialEntropyCompostionAndTemperatureProfiles_dq0.01} shows the entropy and temperature profiles for the narrow gradient. 
In this step, the model is not forced to follow the adiabatic gradient and the outer boundary condition is given by the default photosphere without irradiation.

We consider two cases for the composition profile, one with a wide transition region (transition over $\Delta q=0.1$) and one with a narrow transition region (transition over $\Delta q=0.001$\footnote{The published version contained a typo, stating the width of the narrow gradient as $\Delta q =0.01$. The correct value used in the simulations is $\Delta q=0.001$.}). Both profiles are fixed to the same bulk metallicity of $Z\low{bulk}=0.95$ with an envelope metallicity of $Z\low{env}=0.2$.

\item \label{item:StepIrradiation} 
Using the options \verb|irradiation_flux| we heat the outer parts of the planet over $100\,000$ years sufficiently high such that changing to a different atmosphere model does not lead to convergence issues. We find a flux of $8\times10^6$ erg/s/cm$^2$ and a column depth of 100 g/cm$^2$ a good choice of parameters for our desired equilibrium temperature.

\item \label{item:StepAtmosphere} 
Finally we change the atmosphere setting \verb|atm_option| to 'irradiated\_grey'. The initial model is then evolved for another $10\,000$ years before saving it as a \verb|.mod| file.

\end{enumerate}
This process is repeated for the three initial energy states and the two composition profiles. The initial temperature profiles are shown in the lower panels of Figure \ref{fig:InitialEntropyCompostionAndTemperatureProfiles} for the wide gradient, and in Figure \ref{fig:InitialEntropyCompostionAndTemperatureProfiles_dq0.01} for the narrow gradient. 
For all initial models, we use the thermal conductivity model corresponding to Cond-1.


\section{Radius evolution for the small and large planet}\label{sec:MoreRadiusEvolution}
\begin{figure}[!ht]
    \centering
    \includegraphics[width=1\linewidth]{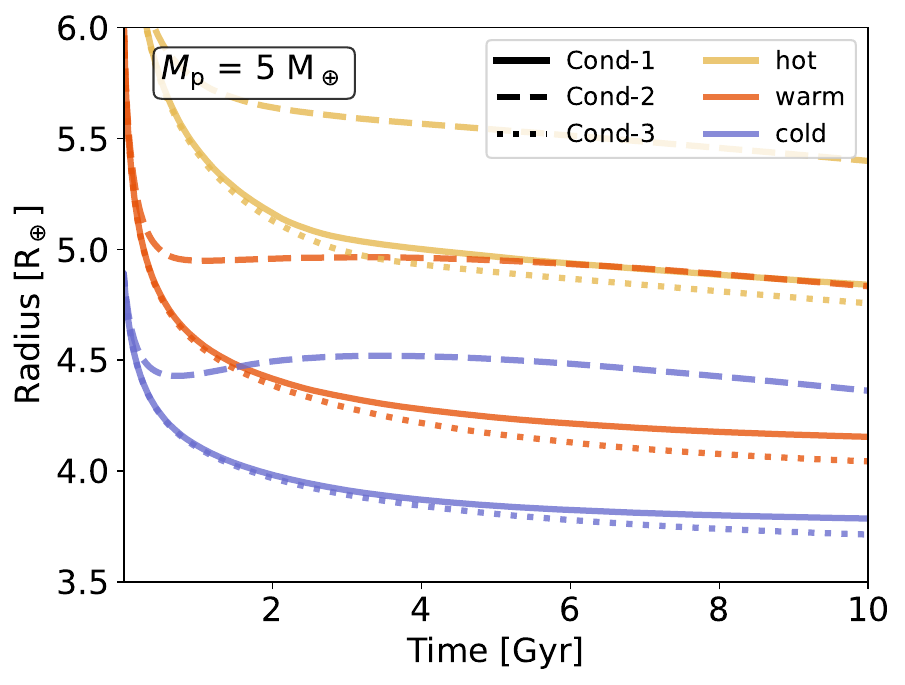}
    \includegraphics[width=1\linewidth]{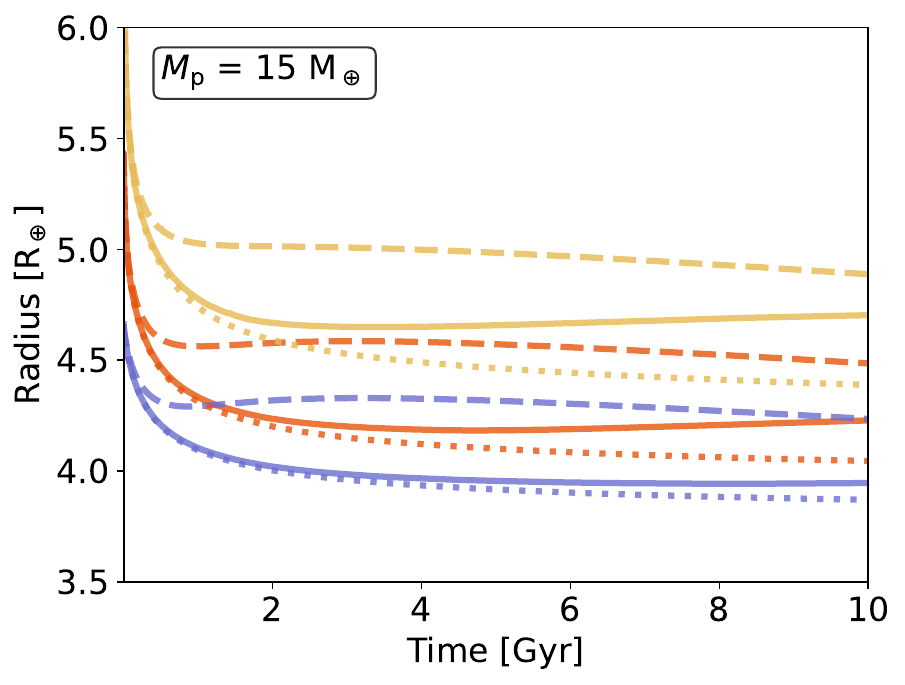}
    \caption{Radius vs.~time for the planets with $M\low{p}=5$~M$_\oplus$ (top panel) and $M\low{p}=15$ M$_\oplus$ (bottom panel). The solid, dashed, and dotted lines correspond to  Cond-1, Cond-2, and Cond-3, respectively. The different colors represent the different initial energy states.}
    \label{fig:RadiusEvolutionM05AndM15}
\end{figure}
For completeness, in this section we show the radius evolution for the three initial entropy cases discussed in Section \ref{sec:CaseComparison}, focusing on planets with $5M_\oplus$ and $15M_\oplus$, as illustrated in \ref{fig:RadiusEvolutionM05AndM15}.

\FloatBarrier
\section{Effects of Cond-1 uncertainties}
\label{sec:Cond1Uncertainty}
\begin{figure}[hbt!]
    \centering
    \includegraphics[width=1\linewidth]{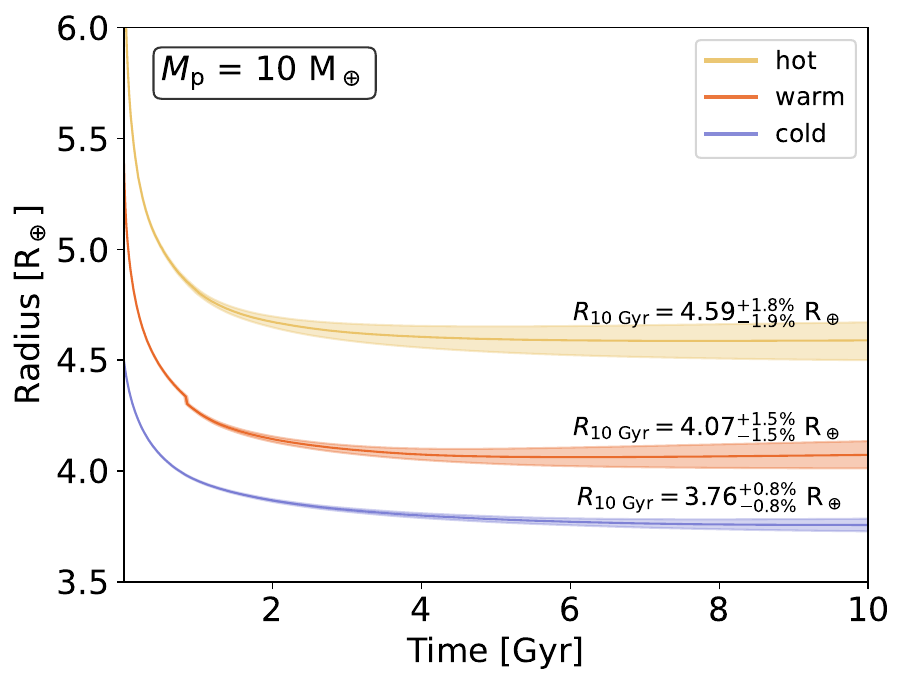}
    \caption{Planetary radius vs.~time for a planet with $M\low{p}=10$~M$_\oplus$. The shaded area shows the uncertainty originating from the statistical error of the conductivity model in Cond-1. The different colors correspond to the  different initial energy states. We annotate the final radius and the  uncertainty associated with the chosen  conductivity model.}
    \label{fig:RadiusEvolutionErrorM10}
\end{figure}
We investigate how the uncertainty in the conductivity inferred with model \mbox{Cond-1} affects the planetary evolution. 
\citet{FrenchRedmer2017} state that the systematic error is expected to be up to $5\%$ and suggest that the statistical error can be estimated based on the scattering of the data points around the fit (see Fig. 4 in French and Redmer 2017). 
Therefore, we take an uncertainty for the vibrational and electronic conductivity of $30\%$ in Cond-1 and scale the conductivities by a factor of $k'=k\,\cdot\,(1\pm0.3)$, where $k$ is the unscaled conductivity of the electronic or vibrational part.
Figure \ref{fig:RadiusEvolutionErrorM10} shows the effect of the uncertainty in conductivity on the radius evolution for the 10 M$_\oplus$ planet. We find small deviations, with the highest being $-2\%$ for the hot 15 M$_\oplus$ planet. Generally,  the deviations are larger for the hotter models because more energy is trapped within the planet.

\section{Radius evolution comparison for Cond-1 and Cond-4}\label{sec:RadiusEvolutionCase4}
\begin{figure}[hbt!]
    \centering
    \includegraphics[width=1\linewidth]{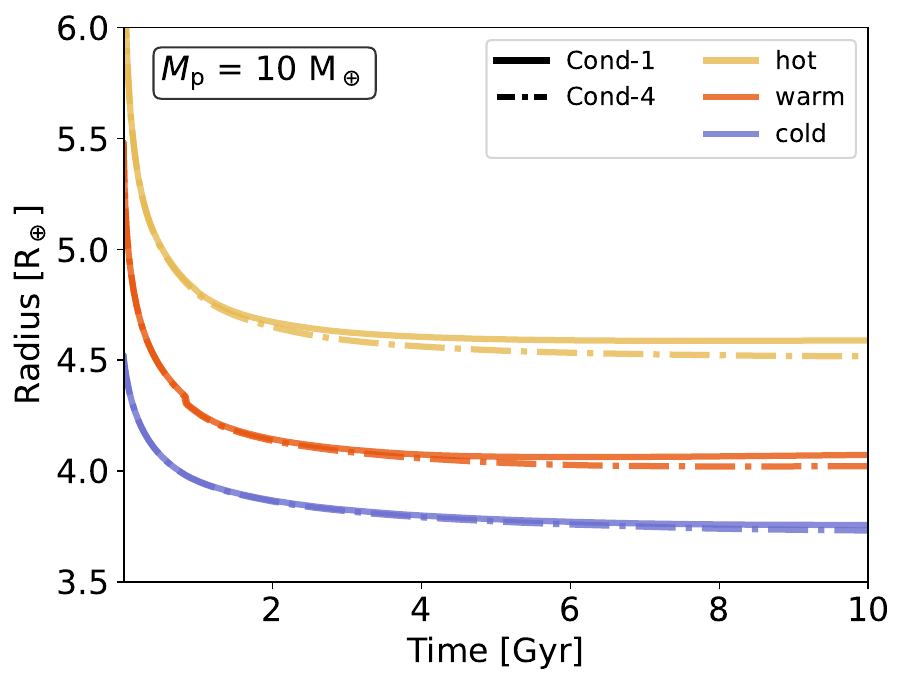}
    \caption{Planetary radius vs.~time for a planet with $M\low{p}=10$~M$_\oplus$. The solid and dashdot line indicates Cond-1 and Cond-4, respectively. The different colors correspond to the  different initial energy states.}
    \label{fig:RadiusEvolutionPhononsM10}
\end{figure}

We compare the radius evolution between Cond-1 and Cond-4 for the $M\low{p}=10$~M$_\oplus$ planet in \ref{fig:RadiusEvolutionPhononsM10}.

\section{The effect of the H-He EoS}\label{sec:RadiusComparisonEoS}
\begin{figure}[hbt!]
    \centering
    \includegraphics[width=1\linewidth]{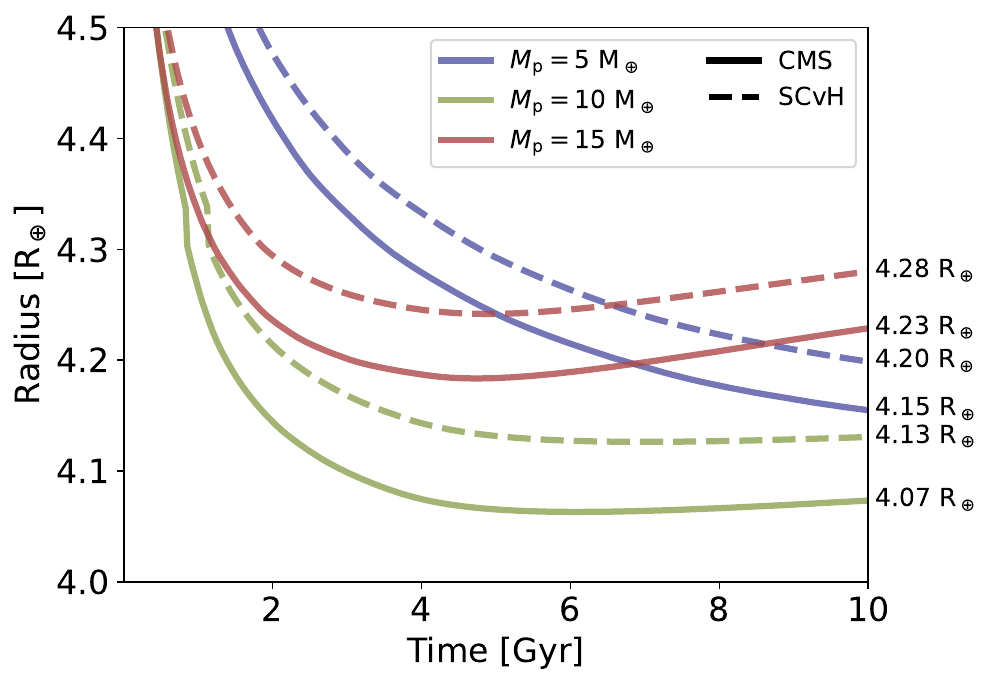}
    \caption{Planetary radius vs.~time for the models with a wide composition gradient and warm initial conditions. The solid and dashed line correspond to the CMS and SCvH EoS for H-He, respectively. The colors, blue, green and red show different planetary masses of $M_\text{p}=$5, 10, and 15M$_\oplus$, respectively. For each simulations we show the radius after $t=10$ Gyr on the right hand side.}
    \label{fig:RadiusEvolutionEoSCompariosn}
\end{figure}

In this study we used the CMS EoS with non-ideal interactions for H-He (see Section \ref{sec:Methods} for details). Here, we present the planetary evolution when using the SCvH EoS for H-He \citep{Saumon+1995} instead of CMS. Figure \ref{fig:RadiusEvolutionEoSCompariosn}  shows the radius evolution for the models with a warm start and a wide composition gradient using both EoS. We find small deviations of $\lesssim2\%$ for the planetary radius after 10 Gyr. Comparing the water EoS using MESA is currently not possible due to the rather complex implementation procedure \citep[e.g.][]{Helled+2025}. 

As noted above, detailed investigations of different EoSs and mixtures that systematically quantify the role of the used EoS and assumed composition are required. This is in particularly timely now as exoplanetary data is improving and theoretical uncertainties can be significant \citep{Mueller+2020b}. 

\end{appendix}
\end{document}